\documentclass[11pt,twoside]{article}

\usepackage{asp2004}
\usepackage{epsf}
\usepackage{psfig}
\usepackage{lscape}

\begin{document}
\title{SOLIS-VSM Solar Vector Magnetograms}

\author{C. J. Henney$^1$, C. U. Keller$^{1,2}$ and J. W. Harvey$^1$}
\affil{$^1$National Solar Observatory, Tucson, Arizona, USA}
\affil{$^2$Sterrenkundig Instituut, Utrecht University, Utrecht, The Netherlands}

\begin{abstract}
The Vector SpectroMagnetograph (VSM) instrument has recorded full-disk 
photospheric vector magnetograms weekly since August 2003 as part of 
the Synoptic Optical Long-term Investigations of the Sun (SOLIS) project. 
After the full deployment of the VSM data processing system, a typical 
observing day will include three Fe~{\footnotesize I} 630.2 nm full-disk 
photospheric vector magnetograms, one full-disk photospheric and 
three Ca~{\footnotesize II} 854.2 nm chromospheric longitudinal magnetograms, 
along with three He~{\footnotesize I} 1083 nm spectroheliograms. The 
photospheric vector magnetograms will be available over the Internet in 
two stages: first, as a quick-look product within minutes of data 
acquisition, and then as a Milne-Eddington inversion product within a day 
of each observation.
\end{abstract}

\section{Introduction}
The VSM is one of three instruments that constitute the SOLIS project. For a 
general overview of the SOLIS instruments see \citet{Keller03}. The VSM instrument 
provides a unique record of solar full-disk vector magnetograms along with 
the high sensitivity photospheric and chromospheric longitudinal magnetograms. 
The VSM began recording daily full-disk magnetograms during August 2003 at a 
temporary site in Tucson, Arizona. In April 2004, the VSM was relocated to Kitt 
Peak and resumed operation in May 2004.

The VSM full-disk parametergrams are constructed from 2048 individual steps
in declination of the projected solar image on the entrance slit. The observed
spectrum lines would be curved due to the relatively large angular deviation 
of 0.57 degrees in the spectrograph. However, the slit is curved with a radius 
of 258.77 mm to compensate for the curvature of the 630.2 nm wavelength spectrum.
The 50-cm aperture VSM utilizes a Ritchey-Chr\'etien optical design. This 
includes a two-lens field corrector to provide nearly equivalent image quality 
over the whole field of view with minimal geometric distortion and equal image 
size for all wavelengths. The VSM utilizes two interim CMOS hybrid cameras 
made by Rockwell Scientific. During the first year of operation, several 
unexpected camera signals were revealed \citep[see][]{Harvey04}. Two of 
the pronounced artifacts, a variable dark level and signal cross-talk at 
a 0.5 percent level, required significant modification to the data 
reduction pipeline.

%%%%%%%%%%%%%%%%%%%%%%%%%%%%%%%%%%%%%%%%%%%%%%%%%%%%%%%%%%%%%%%%%%
%%
\begin{figure}[!t]
  \plotone{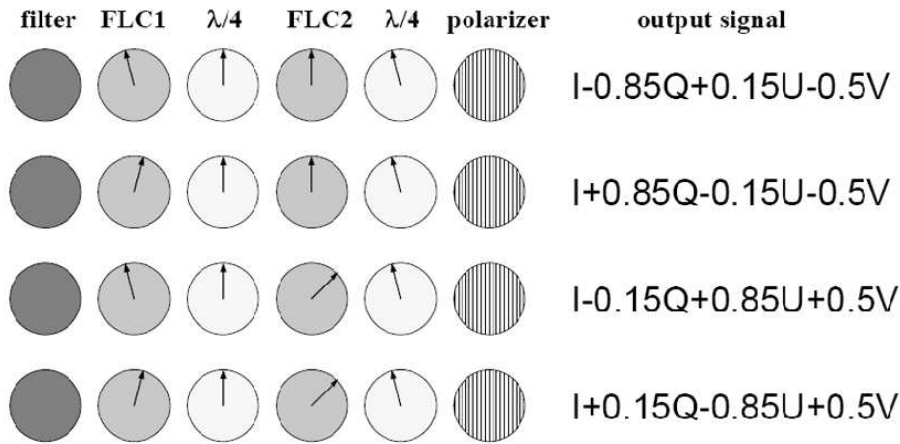}
  \caption{Illustration of the VSM optical sequence for each of the
four modulation states to measure the four Stokes parameters (see text).}
\end{figure}
%%
%%%%%%%%%%%%%%%%%%%%%%%%%%%%%%%%%%%%%%%%%%%%%%%%%%%%%%%%%%%%%%%%%%

\subsection{Polarization Calibration}
The optical configurations for the VSM polarization modulation measurements 
used to produce vector magnetograms are depicted in Figure~1. VSM vector 
magnetograms are obtained using two half-wave (at 630 nm) ferroelectric liquid 
crystal (FLC) retarders. In front of the FLCs, off-band wavelengths are filtered 
out. In between and after the FLCs there are polymer quarter-wave retarders
with fixed optic axes (see Figure~1). The arrows in the figure portray the 
relative orientations of each optic axis, where the corresponding output signals 
for each configuration are listed in the right-hand column.

Detailed quantitative calibration of the early VSM vector observations is nearly 
complete. Calibration measurements are obtained by observing at disk center while 
frames are accumulated with an additional polarizer and quarter-wave plate in the
beam before the first FLC in Figure~1. For full-Stokes calibration, the signal 
matrices are estimated by positioning the polarizer and quarter-wave plate 
at 64 angle-positions with respect to each other and the entrance slit. 
In addition, the retardation and relative orientation of the fast axis of 
the calibration retarder are estimated.

%%%%%%%%%%%%%%%%%%%%%%%%%%%%%%%%%%%%%%%%%%%%%%%%%%%%%%%%%%%%%%%%%%
%%
\begin{figure}[!p]
  \plotone{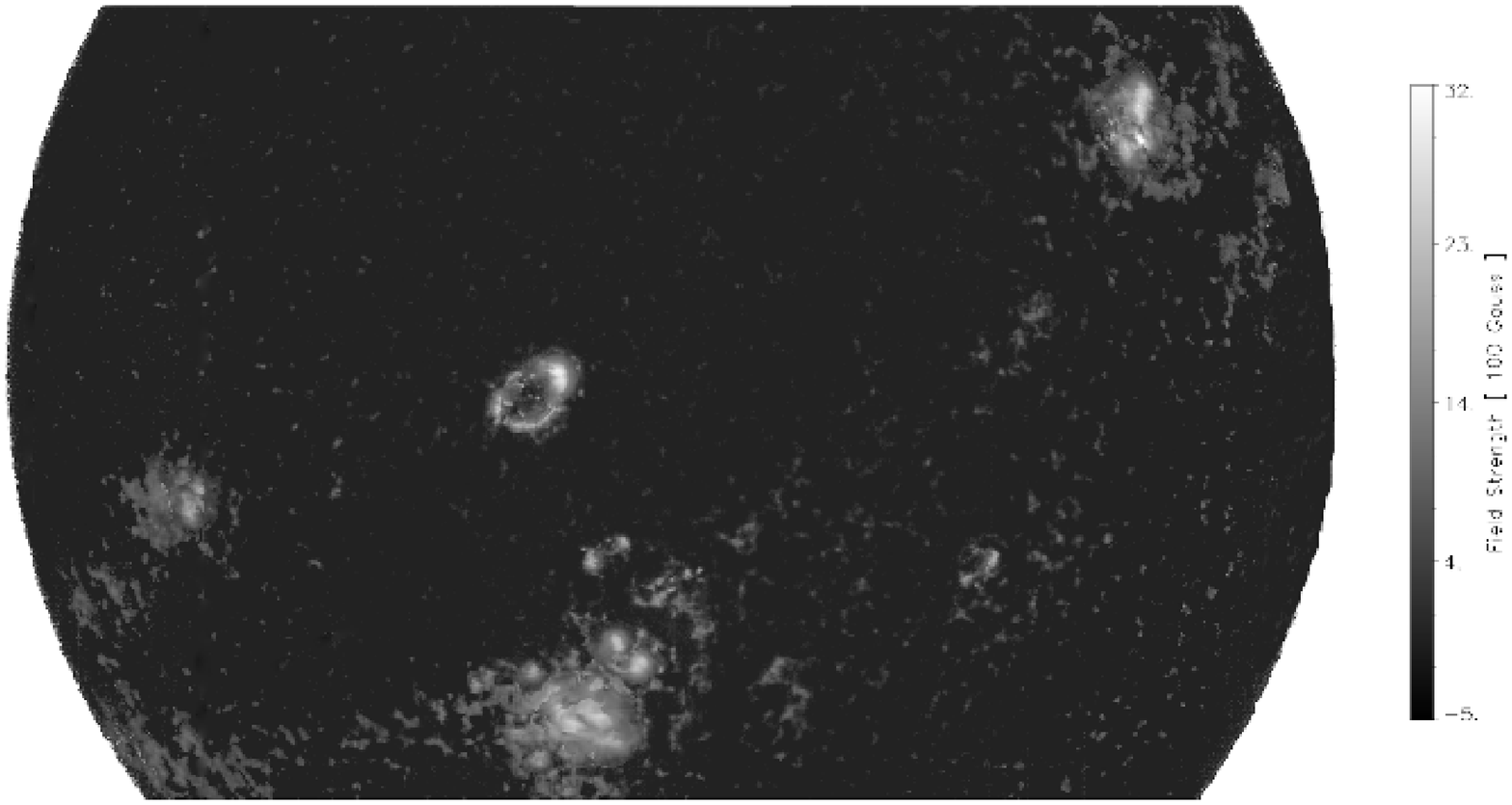}
  \plotone{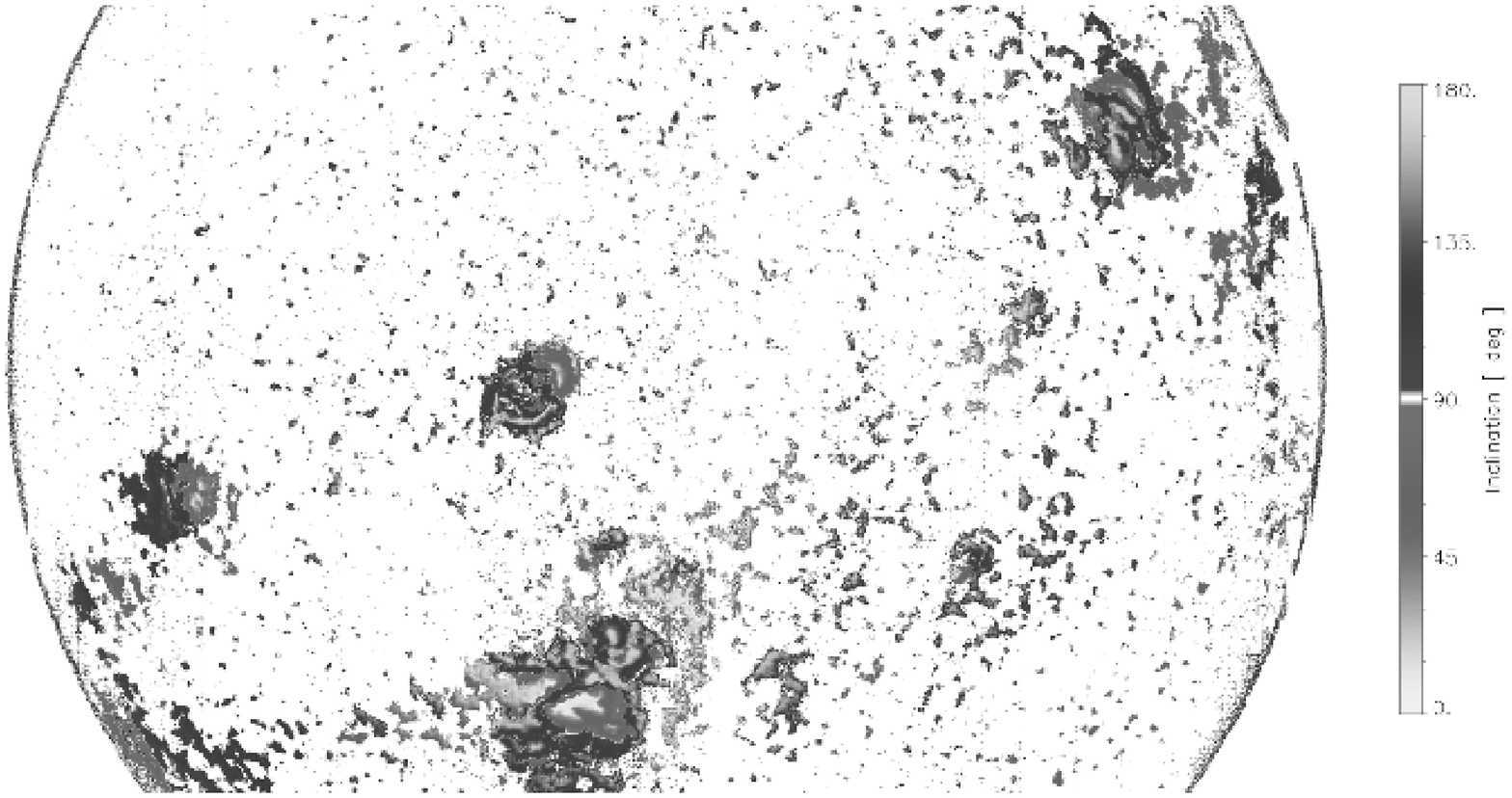}
  \plotone{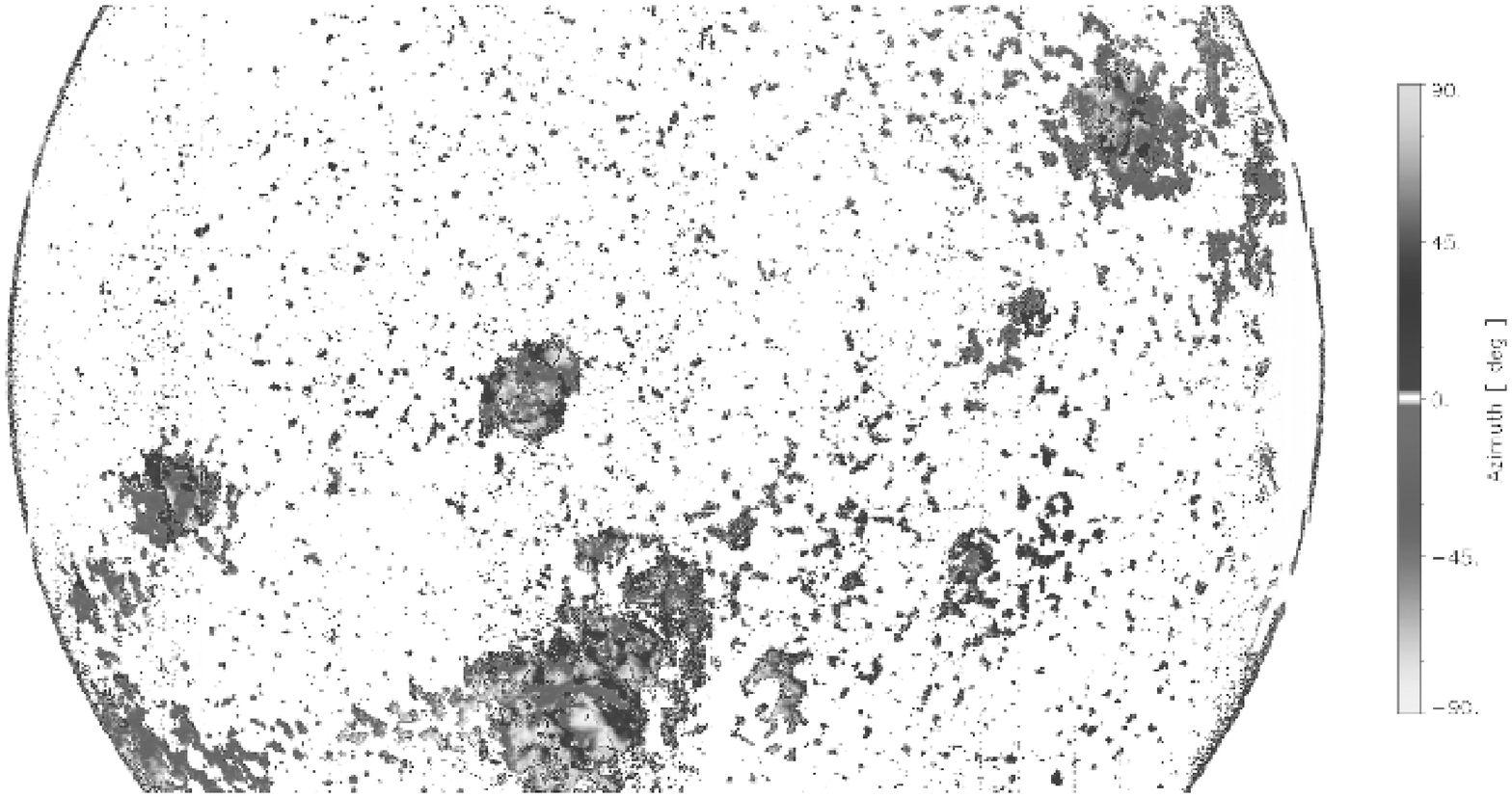}
  \caption{Inferred magnetic field strength (top), inclination (middle), and 
azimuth (bottom) from a SOLIS-VSM area-scan observation on 27/10/2003.}
\end{figure}
%%
%%%%%%%%%%%%%%%%%%%%%%%%%%%%%%%%%%%%%%%%%%%%%%%%%%%%%%%%%%%%%%%%%%

\section{VSM Data Processing}
The SOLIS Data Handling System (DHS) has been designed to use pools
of distributed data reduction processes that are allocated to different 
observations on 10 dual-CPU Linux boxes \citep[see][]{Wamp02,Jones02}.
The SOLIS data transmission and storage resources are such that the reduction of
VSM data is performed at the observing site on Kitt Peak. Reduced data products 
are transmitted via a 45 Mb/s link from Kitt Peak to the National Solar 
Observatory's digital archive in Tucson. Currently, the raw data is archived
for planned reprocessing during 2006. However, during nominal operation the
profile data will be deleted after 24 hours. 

\subsection{Stokes Inversion}
During a typical observing day three full-disk photospheric vector magnetograms
will be available over the Internet in two stages: first, as 
a ``quick-look'' (QL) product within 10 minutes of data acquisition, and then as 
a Milne-Eddington (ME) inversion product within 24 hours of each observation. 
The quick-look parameters
include estimates of the magnetic field strength, inclination, and azimuth.
The QL inclination and azimuth determination is estimated following \citet{Auer77}. 
Whereas, the magnetic field strength is approximated by using the estimated 
inclination with the measured line-of-sight (LOS) field strength
and assumption that the fill-factor is 1. The LOS field strength is 
estimated using the center-of-gravity method \citep{Rees79}. 
The high-precision vector parameters are determined with the High Altitude 
Observatory ME inversion technique implemented by \citet{Sku87}. Example 
ME parameter images from a VSM area-scan observed 27 October 2003 are 
displayed in Figure~2. Initially, the vector data products will not include 
azimuth disambiguation until suitable 
algorithms have been developed. However, the flexible design of the VSM data 
handling system can incorporate azimuth disambiguation and additional future 
improvements under consideration (e.g. principal component analysis). 

\section{VSM Data Products}
New data products from VSM observations since August 2003 are expected to 
become available during 2006. These products include full-disk vector 
magnetograms, Dopplergrams, equivalent width and line depth images. 
Carrington rotation and daily synoptic maps are now available for the 
photospheric magnetograms. Daily coronal hole estimate images derived from 
VSM data, along with coronal hole synoptic maps, are available. 
The VSM synoptic data products are available on the NSO-SOLIS web 
site at: http://solis.nso.edu/.

\section{Acknowledgments}
The authors acknowledge many years of dedicated effort by the SOLIS team.
We are also grateful to K.D. Leka for numerous helpful discussions. SOLIS VSM 
data used here are produced cooperatively by NSF/NSO and NASA/GSFC. 
The National Solar Observatory is operated by AURA, Inc. under a 
cooperative agreement with the National Science Foundation.


\begin{thebibliography}{}

\bibitem[Auer, Heasley \& House(1977)]{Auer77}
Auer, L. H., Heasley, J. N., \& House, L. L. 1977, Solar Physics, 55, 47

\bibitem[Harvey et al.(2004)]{Harvey04}
Harvey, J., Keller, C., Cole, L., Tucker, R., \& Jaksha, D. 2004, 
Proc. SPIE, 5171, 258

\bibitem[Jones et al.(2002)]{Jones02}
Jones, H. P., Harvey, J. W., Henney, C. J., Hill, F., \& Keller, C. U. 2002,
Proc. IAU Colloq. 188, ESA SP-505, 15

\bibitem[Keller, Harvey, \& Giampapa(2003)]{Keller03}
Keller, C. U., Harvey, J. W. \& Giampapa, M. S. 2003, Proc. SPIE, 4853, 194

\bibitem[Rees \& Semel(1979)]{Rees79}
Rees, D. E. \& Semel, M. D. 1979, A\&A, 74, 1

\bibitem[Skumanich \& Lites(1987)]{Sku87}
Skumanich, A. \& Lites, B. W. 1987, ApJ, 322, 473

\bibitem[Wampler(2002)]{Wamp02}
Wampler, S. 2002, Proc. SPIE, 4848, 85

\end{thebibliography}
\end{document}